# Exponential Spectral Risk Measures

By

Kevin Dowd and John Cotter[*]

Abstract

Spectral risk measures are attractive risk measures as they allow the user to obtain risk measures that reflect their subjective risk-aversion. This paper examines spectral risk measures based on an exponential utility function, and finds that these risk measures have nice intuitive properties. It also discusses how they can be estimated using numerical quadrature methods, and how confidence intervals for them can be estimated using a parametric bootstrap. Illustrative results suggest that estimated exponential spectral risk measures obtained using such methods are quite precise in the presence of normally distributed losses.

March 20, 2007

[*] Kevin Dowd is at the Centre for Risk and Insurance Studies, Nottingham University Business School, Jubilee Campus, Nottingham NG8 1BB, UK; email: Kevin.Dowd@nottingham.ac.uk. John Cotter is at the Centre for Financial Markets, School of Business, University College Dublin, Carysfort Avenue, Blackrock, Co. Dublin, Ireland; email: john.cotter@ucd.ie. The authors would like to thank Carlo Acerbi and Dirk Tasche for fruitful conversations on the subject. Cotter's contribution to the study has been supported by a University College Dublin School of Business research grant.

# 1. Introduction

One of the most interesting and potentially most promising recent developments in the financial risk area has been the theory of spectral financial risk measures (SRMs), recently proposed by Acerbi (2002, 2004). SRMs belong to the family of coherent risk measures proposed by Artzner *et alia* (1997, 1999), and therefore possess the highly desirable property of subadditivity[1]. It is also well-known by now that the most widely used risk measure, the Value-at-risk (VaR), is not subadditive, and the work by Artnzer *et alia* and Acerbi has shown that many (if not most) of the inadequacies of VaR as a risk measure can be traced to its non-subadditivity.

One of the nice features of SRMs is that they relate the risk measure itself to the user's subjective risk-aversion – in effect, the spectral risk measure is a weighted average of the quantiles of a loss distribution, the weights of which depend on the user's risk-aversion function. Spectral risk measures enable us to link the risk measure to the user's attitude towards risk, the underlying objective being to ensure that if a user is more risk averse, other things being equal, then that user should face a higher subjective risk, which is what the SRM measures. This means, for example, that two different investors might have the same portfolios and share the same set of forecasts, but their subjective risk measures will still be different if one of them is more risk averse than the other.

In principle, SRMs can be applied to any problems involving risky decision making. Amongst many other possible applications, Acerbi (2004) suggests that they can be used to set capital requirements or obtain optimal risk-expected return tradeoffs in portfolio analysis, and Cotter and Dowd (2006) suggest that SRMs could be used by futures clearinghouses to set margin requirements that reflect their corporate risk appetites.

This paper investigates SRMs further. In particular, it focuses on SRMs based on an underlying exponential utility function. Our objective is two-fold.

---

[1] More formally, if $\rho(.)$ is a measure of risk, and *A* and *B* are any two positions, subadditivity means that $\rho(A+B) \leq \rho(A) + \rho(B)$. Subadditivity is a crucial condition because it ensures that our risks do not increase overall when we put them together. As Acerbi and others have pointed out, any risk measure that does not satisfy subadditivity has no real claim to be regarded as a 'respectable' risk measure at all (see, e.g., Acerbi (2004, p. 150)).

First, we seek to establish some of the properties of SRMs to see how intuitive and 'well-behaved' they might be. Our second objective is computational: we discuss how these SRMs might be estimated and also discuss how we might estimate confidence intervals for them.

The article is organised as follows. Section 2 sets out the essence of Acerbi's theory of spectral risk measures. Section 3 discusses SRMs based on exponential utility functions. Section 4 discusses the estimation of SRMs and section 5 discusses the estimation of confidence intervals for them. Section 6 concludes.

**2. Spectral Risk Measures**

Following Acerbi (2004), consider a risk measure $M_\phi$ defined by:

$$(1) \qquad M_\phi = \int_0^1 \phi(p) q_p \, dp$$

where $q_p$ is the $p$ loss quantile and $\phi(p)$ is a user-defined weighting aversion function with weights defined over $p$, where $p$ is a continuous range of cumulative probabilities $p \in [0,1]$. We can think of $M_\phi$ as the class of quantile-based risk measures, where each individual risk measure is defined by its own particular weighting function.

Two well-known members of this class are the VaR and the Expected Shortfall (ES). The VaR at the $\alpha$ confidence level is:

$$(2) \qquad VaR_\alpha = q_\alpha$$

The VaR places all its weight on a single quantile that corresponds to a chosen confidence level, and places no weight on any others, i.e., with the VaR risk measure, $\phi(p)$ takes the degenerate form of a Dirac delta function that gives the outcome $p = \alpha$ an infinite weight and gives every other outcome a zero weight.

For its part, the ES at the confidence level $\alpha$ is the average of the worst $1-\alpha$ of losses and (in the case of a continuous loss distribution) is:

(3) $$ES_\alpha = \frac{1}{1-\alpha} \int_\alpha^1 q_p \, dp$$

With the ES, $\phi(p)$ gives tail quantiles a fixed weight of $\frac{1}{1-\alpha}$ and gives non-tail quantiles weight of zero.

A drawback with both of these risk measures is that they inconsistent with risk aversion in the traditional sense. This can be illustrated in the context of the theory of lower partial moments (see Bawa (1975), Fishburn (1977) and Grootveld and Hallerbach (2004)). Given a set of returns $r$ and a target return $r*$, the lower partial moment of order $k \geq 0$ around $r*$ is equal to $E\{[\max(0, r*-r]^k\}$. The parameter $k$ reflects the user's degree of risk aversion, and the user is risk-averse if $k > 1$, risk-neutral if $k = 1$ and risk-loving if $0 < k < 1$. It can then be shown that the VaR is a preferred risk measure only if $k = 0$, i.e., the VaR is our preferred risk measure only if we are very risk-loving! The ES would be our preferred risk measure if $k = 1$, and this tells us that the ES is our preferred risk measure only if the user is risk-neutral between better and worse tail outcomes.

A user who is risk averse might prefer to work with a risk measure that take account of his/her risk aversion, and this takes us to the class of spectral risk measures (SRMs). In loose terms, an SRM is a quantile-based risk measure that takes the form of (1) where $\phi(p)$ reflects the user's degree of risk aversion. More precisely, we can consider SRMs as the subset of $M_\phi$ that satisfy the following properties of positivity, normalisation and increasingness due originally to Acerbi:[2]

---

[2] See Acerbi (2002, 2004). However, it is worth pointing out that he deals with a distribution in which profit outcomes have a positive sign, whereas we deal with a distribution in which loss outcomes have a positive sign. His first condition is therefore a negativity condition, whereas ours is a positivity condition, but this difference is only superficial and there is no substantial difference between his conditions and ours.

*1. Positivity*: $\phi(p) \geq 0$.

*2. Normalisation*: $\int_0^1 \phi(p)dp = 1$.

*3. Increasingness*: $\phi'(p) \geq 0$.

The first coherent condition requires that the weights are weakly positive and the second requires that the probability-weighted weights should sum to 1, but the key condition is the third one. This condition is a direct reflection of risk-aversion, and requires that the weights attached to higher losses should be no less than the weights attached to lower losses. Typically, we would also expect the weight $\phi(p)$ to rise with $p$.[3] In a 'well-defined' case, we would expect the weights to rise smoothly, and the more risk-averse the user, the more rapidly we would expect the weights to rise.

A risk measure that satisfies these properties is attractive not only because it takes account of user risk-aversion, but also because such a risk measure is known to be coherent.[4]

However, there still remains the question of how to specify $\phi(p)$, and perhaps the most natural way to obtain $\phi(p)$ is from the user's utility function[5].

## 3. Exponential Spectral Risk Measures

This requires us to specify the utility function, and a common choice is the exponential utility function. This utility function is defined conditional on a single parameter, the coefficient of absolute risk aversion. The exponential utility function is defined as follows over outcomes *x*:

(4) $$U(x) = -e^{-ax}$$

---

[3] The conditions set out allow for the degenerate limiting case where the weights are flat for all p values, and such a situation implies risk-neutrality and is therefore inconsistent with risk-aversion. However, we shall rule out this limiting case by imposing the additional (and in the circumstances very reasonable) condition that $\phi(p)$ must rise over at least point as *p* increases from 0 to 1.

[4] The coherence of SRMs follows from Acerbi (2004, Proposition 3.4).

[5] See also Bersimas *et alia* (2004).

where $a > 0$ is the Arrow-Pratt coefficient of absolute risk aversion (ARA). The coefficients of absolute and relative risk aversion are:

(5a) $$R_A(x) = -\frac{U''(x)}{U'(x)} = a$$

(5b) $$R_R(x) = -\frac{xU''(x)}{U'(x)} = xa$$

We now set

(6) $$\phi(p) = \lambda e^{-a(1-p)}$$

where $\lambda$ is an unknown positive constant. This clearly satisfies properties 1 and 3, and we can easily show (by integrating $\phi(p)$ from 0 to 1, setting the integral to 1 and solving for $\lambda$) that it satisfies property 2 if we set

(7) $$\lambda = \frac{a}{1-e^{-a}}$$

Hence, substituting (7) into (6) gives us the exponential weighting function (or risk-aversion function) corresponding to (4):[6]

(8) $$\phi(p) = \frac{ae^{-a(1-p)}}{1-e^{-a}}$$

This risk-aversion function is illustrated in Figure 1 for two alternative values of the ARA coefficient, $a$. Observe that this weighting function has a nice shape: for the higher $p$ values associated with higher losses, we get bigger weights for greater degrees of risk-aversion. In addition, as $p$ rises, the rate of increase of $\phi(p)$ rises with the degree of risk-aversion.

---

[6] Strictly speaking, Acerbi's proposition 3.19 in Acerbi (2004, p. 182) defines his weighting function in terms of a parameter $\gamma > 0$, but his weighting function and (7) are equivalent subject to the proviso that $\gamma = 1/a$.

**Insert Figure 1 here**

The SRM based on this risk-aversion function, the exponential SRM, is then found by substituting (8) into (1), viz.:[7]

(9) $$M_\phi = \int_0^1 \phi(p) q_p \, dp = M_\phi = \frac{a}{1-e^{-a}} \int_0^1 e^{-a(1-p)} q_p \, dp$$

We also find that the risk measure itself rises with the degree of risk-aversion, and some illustrative results are given in Table 1. For example, if losses are distributed as standard normal and we set $a = 5$, then the spectral risk measure is 1.0816. But if we increase $a$ to 25, the measure rises to 1.9549: the greater the risk-aversion, the higher the exponential spectral risk measure.

**Insert Table 1 here**

The relationship of the exponential SRM and the coefficient of absolute risk aversion is illustrated further in Figure 2. We can see that the risk measure rises smoothly as the coefficient of risk aversion increases, as we would expect.

**Insert Figure 2 here**

These results suggest that ESRMs are 'well-behaved' and have nice intuitive properties.

**4. Estimating Spectral Risk Measures**

---

[7] Estimates of (9) were obtained using Simpson's rule numerical quadrature with $p$ divided into $N$=10,000,001 'slices'. The calculations were carried out using the CompEcon functions in MATLAB given in Miranda and Fackler (2002). As the results in section 4 below suggest, this value of $N$ suffices to us effectively exact estimates of SRMs. We use $n$=10,000,001 rather than the more obvious $n$=10,000,000 because the Simpson's rule algorithm requires $n$ to be odd.

The equation for the SRM, (9), indicates that solving for the SRM involves integration. Special cases aside, this integration would need to be carried out numerically rather than analytically. This means that the estimation of SRMs requires a suitable numerical integration or numerical quadrature method. Such methods estimate the integral from a numerical estimate of its discretised equivalent in which $p$ is broken into a large number of $N$ 'slices'.[8] This, in turn, raises the question of how different quadrature methods compare. Furthermore, since quadrature methods depend on a specified value of $N$, it also raises the issue of how estimates might depend on the value of $N$.

To investigate these issues further, Figure 3 provides some plots showing how estimates of standard normal SRMs vary with different quadrature methods and different values of $N$. These plots are based on an illustrative ARA coefficient equal to 5, but we get similar plots for other values of this coefficient. The methods examined are those based on the trapezoidal rule, Simpson's rule, and Niederreiter and Weyl quasi-Monte Carlo. As we might expect, all four quadrature methods give estimates that converge on their true values as $N$ gets larger. However, the trapezoidal and Simpson's rule methods produce estimates that converge smoothly as $N$ gets larger, whereas the two quasi-Monte Carlo methods produce estimates that converge more erratically as $N$ gets larger. The plots also suggest that the first two methods are usually more accurate for any given value of $N$, and that the method based on Simpon's rule is marginally better than that based on the trapezoidal rule.

**Insert Figure 3 here**

In addition, these plots show that all methods produce estimates of SRMs that have a small downward bias. If we wish to get accurate estimates of SRMs, it is therefore important to choose $N$ value large enough to make this bias negligible. To investigate further, Table 2 reports results for the accuracy and calculation times of the same SRM estimated using Simpson's rule and different values of $N$. This shows that, for $N$=1,001, we get an estimate with an error of -1.55% and this takes 0.0027 seconds to calculate using the latest version (version 2007a) of MATLAB using a Pentium 4 desktop computer. For $N$=10,001, the error is -0.18% and this estimate takes 0.0096 seconds to calculate. The accuracy and calculation times increase as $N$ gets larger, and when $N$=10,000,001, we get an error of -0.00% and a calculation time of a little over 8 seconds. Bearing in mind that real-world risk measurement is subject to many different sources of error that are beyond the control of the risk manager, it is pointless to go for spuriously accurate estimates that ignore these other sources of error. We would therefore suggest that a value of $N$=10,001 is for practical purposes accurate enough for the needs of risk managers in the real world.[9]

---

[8] For more on these methods, see, e.g., Borse (1997, chapter 7), Miranda and Fackler (2002, chapter 5) or Kreyszig (1999, chapter 17).

**Insert Table 2 here**

**5. Estimating Confidence Intervals for Spectral Risk Measures**

Given that our estimates are prone to many sources of error, it is good practice to estimate some precision metrics to go with our estimated risk measures, and perhaps the best such metrics are estimates of their confidence intervals.

We can easily obtain such estimates using a parametric bootstrap, and this can be implemented using the following procedure:

- Given $N$, for each of $b$ bootstrap trials, we simulate a set of $N$ loss values from our assumed distribution, in this case a standard normal. We order these simulated losses from lowest to highest to obtain a set of simulated quantiles, $\tilde{q}_p$. We then apply our chosen quadrature method to (9) with $q_p$ replaced by $\tilde{q}_p$ to obtain a bootstrap estimate of the SRM.

- We repeat this step $b$ times and obtain an estimate of the confidence interval from the distribution of simulated SRM estimates.

---

[9] However, beware the small print: these results are obtained assuming that losses are normal and were also obtained using a good numerical integration routine. In practice, risk managers would be well advised to check the accuracy of their numerical integration routines before using them and also check the extent to which similar findings hold or do not hold for the distributions they are dealing with.

Some illustrative results are shown in Table 3. This shows estimates of the 90% confidence intervals for values of the ARA coefficient equal to 5 and 100. The Table also shows estimates of the standardised confidence intervals, which are equal to the 90% confidence intervals with the bounds divided by the mean of the bootstrap SRM estimates. All results are based on $N$=10,001 and $b$=1000. We can see that for ARA=5, the 90% confidence is [1.0591, 1.1012] and its equivalent for ARA=100 is [2.4075, 2.5380]. For ARA=5, the standardised interval is [0.9805, 1.0195] and for ARA=100 the standardised interval is [0.9739, 1.0267]. It is striking how narrow these intervals are: for ARA=5, the width of the interval is only 3% of the estimated mean SRM, and for ARA=100, the width is only a little over 5% of the estimated mean SRM. These narrow confidence intervals indicate that the SRM estimates are very precise.

**Insert Table 3 here**

**6. Conclusions**

This paper has examined spectral risk measures based on an exponential utility function. We find that the exponential utility function leads to risk-aversion functions and spectral risk measures with intuitive and nicely behaved properties. These exponential SRMs are easy to estimate using numerical quadrature methods and accurate estimates can be obtained very quickly in real time. It is also easy to estimate confidence intervals for these SRMs using a parametric bootstrap. Illustrative results suggest that these confidence intervals are surprisingly narrow, and this indicates that SRM estimates are quite precise. Of course, the results presented here are based on an assumed normal distribution, and further work is needed to establish results for other distributions.[10]

---

[10] However, a start has already been made: for example, Acerbi (2004, pp. 202-205) provides some results for lognormal and power-law distributions, and Cotter and Dowd (2006) provide results for Generalised Pareto distributions.

# References

Acerbi, C., (2002) "Spectral Measures of Risk: A Coherent Representation of Subjective Risk Aversion." *Journal of Banking and Finance*, 26, 1505-1518.

Acerbi, C., (2004) "Coherent Representations of Subjective Risk Aversion." Pp. 147-207 in G. Szegö (Ed.) *Risk Measures for the 21$^{st}$ Century*, New York: Wiley.

Artzner, P., F. Delbaen, J.-M. Eber, and D. Heath, (1997) "Thinking coherently." *Risk*, 10 (November), 68-71.

Artzner, P., F. Delbaen, J.-M. Eber, and D. Heath, (1999) "Coherent Measures of Risk." *Mathematical Finance*, 9, 203-228.

Bawa, V. S., (1975) "Optimal Rules for Ordering Uncertain Prospects." *Journal of Financial Economics*, 2, 95-121.

Borse, G. J. (1997) *Numerical Methods with MATLAB: A Resource for Scientists and Engineers*. Boston: PWS Publishing Company.

Bertsimas, D., G. J. Lauprete, and A. Samarov, (2004) "Shortfall as a Risk Measure: Properties, Optimization and Applications." *Journal of Economic Dynamics and Control*, 28, 1353 – 1381.

Cotter, J., and K. Dowd, (2006) "Extreme Spectral Risk Measures: An Application to Futures Clearinghouse Margin Requirements." *Journal of Banking and Finance*, 30, 3469-3485.

Dowd, K., (2005) *Measuring Market Risk*. Second edition, Chichester: John Wiley and Sons.

Fishburn, P. C., (1977) Mean-Risk Analysis with Risk Associated with Below-Target Returns." *American Economic Review*, 67, 116-126.

Grootveld, H., and W. G. Hallerbach, (2004) "Upgrading Value-at-Risk from Diagnostic Metric to Decision Variable: A Wise Thing to Do?" Pp. 33-50 in G. Szegö (Ed.) *Risk Measures for the 21$^{st}$ Century*, Wiley, New York.

Kreyszig, E. (1999) *Advanced Engineering Mathematics*. 8$^{th}$ edition. Wiley, New York.

Miranda, M. J., and P. L. Fackler, (2002) *Applied Computational Economics and Finance*. Cambridge, MA and London: MIT Press.

**FIGURES**

**Figure 1: Exponential Risk Aversion Functions**

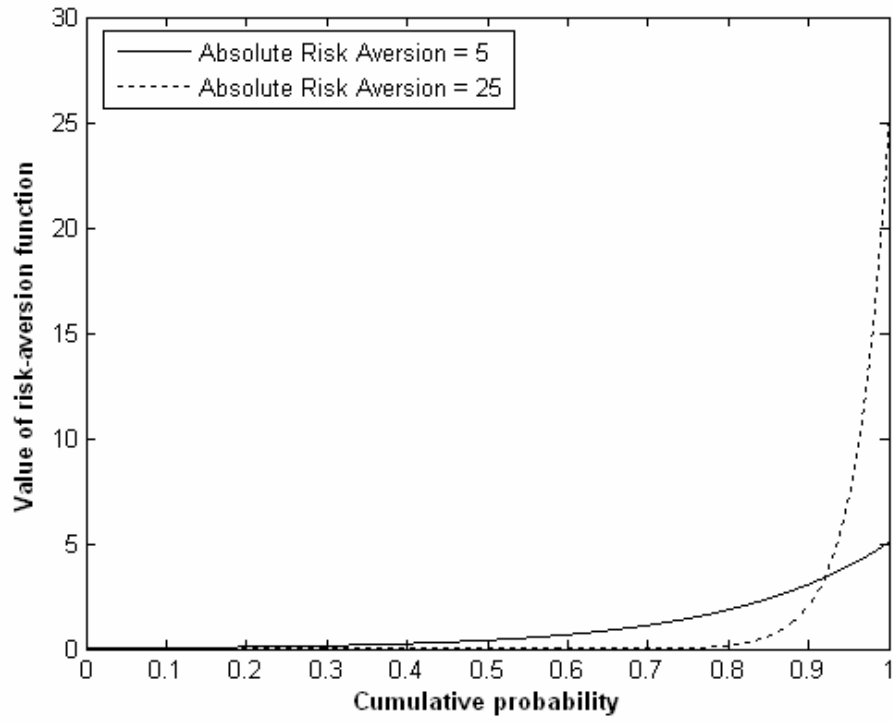

Notes: Weights are based on the exponential risk-aversion function (8).

**Figure 2: Plot of Exponential Spectral Risk Measure Against the Coefficient of Absolute Risk Aversion: Standard Normal Loss Distribution**

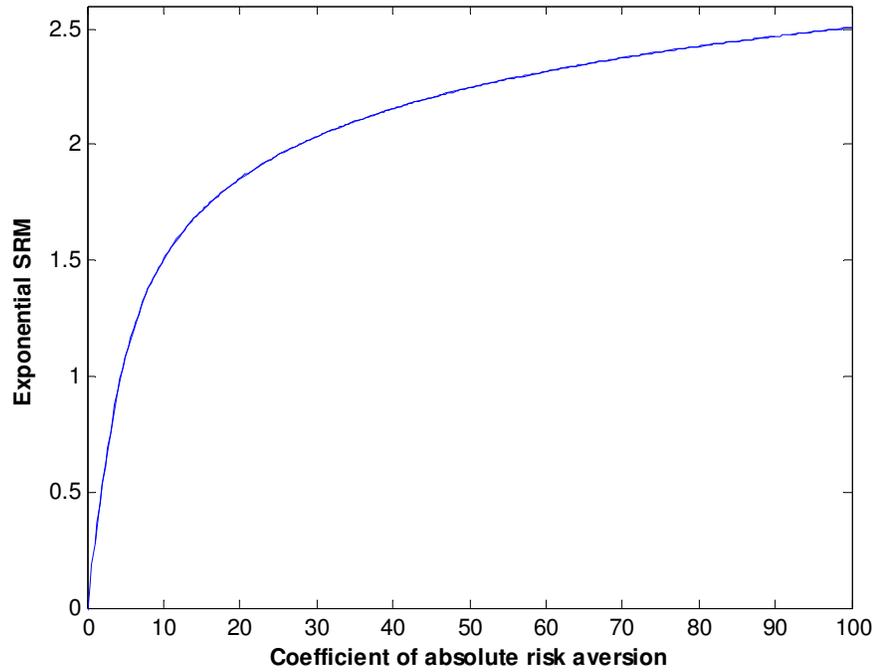

Notes: As per Table 1.

**Figure 3: Estimates of Standard Normal Spectral Risk Measure for Various Numerical Quadrature Methods and Values of *N*: Coefficient of Absolute Risk Aversion = 5**

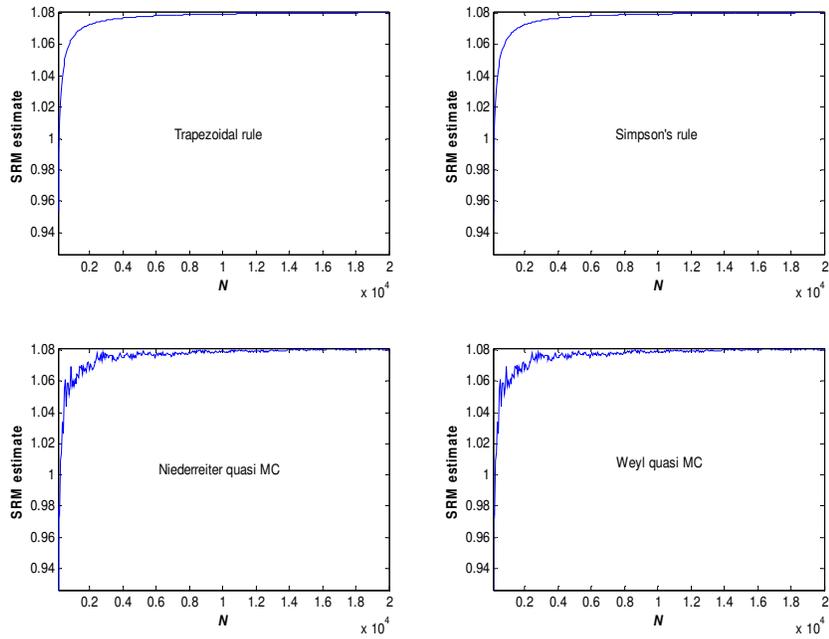

Notes: As per Table 1. Each chart gives plots of the estimated SRM predicated on ARA=5 for each of four numerical quadrature methods: the trapezoidal rule, Simpson's rule, the Niederreiter quasi Monte-Carlo and Weyl quasi-Monte Carlo. Estimates are of (9) obtained for values of *N* ranging from 100 to 20000.

**Table 1: Values of Exponential Spectral Risk Measure with Standard Normal Losses**

| Coefficient of Absolute Risk Aversion | Exponential Spectral Risk Measure |
|---|---|
| 1 | 0.2781 |
| 5 | 1.0816 |
| 25 | 1.9549 |
| 100 | 2.5055 |

Notes: Estimates are of (9) obtained using Simpson's rule numerical quadrature with $p$ divided into $N$=10,000,001 'slices'. The calculations were carried out using the Miranda-Fackler (2002) CompEcon functions in the 2007a version of MATLAB on a Pentium 4 desktop computer.

**Table 2: Percentage Errors and Calculation Times for Estimates of Standard Normal Exponential Spectral Risk Measure Obtained Using Simpson's Rule: Coefficient of Absolute Risk Aversion = 5**

| N | Percentage Error | Calculation Time (secs) |
|---|---|---|
| 1,001 | -1.55% | 0.0027 |
| 10,001 | -0.18% | 0.0096 |
| 100,001 | -0.03% | 0.0818 |
| 1,000,001 | -0.01% | 0.8689 |
| 10,000,001 | 0.00% | 8.3775 |

Note: As per the Notes to Table 1 but with the specified value of $N$.

**Table 3: Illustrative Estimates of 90% Confidence Intervals for Spectral Risk Measures**

| $ARA = 5$ | $ARA = 100$ |
|---|---|
| *90% Confidence Interval* | |
| [1.0591, 1.1012] | [2.4075, 2.5380] |
| *Standardized 90% Confidence Interval* | |
| [0.9805, 1.0195] | [0.9739, 1.0267] |

Note: Estimates are of (9) obtained using Simpson's rule numerical quadrature with $N$=10,001 and 1000 resamples from a parametric bootstrap. The SRM estimates were obtained using the Miranda-Fackler (2002) CompEcon functions in MATLAB on a Pentium 4 desktop computer. The underlying loss distribution is assumed to be standard normal. Each confidence interval took just over 14 seconds to calculate. The standardised 90% confidence intervals are equal to the associated 90% confidence interval with each bound divided by the mean of the bootstrap SRM estimates.